\documentclass[aps,superscriptaddress,showpacs,
 reprint]{revtex4-1} 
 \usepackage[utf8]{inputenc}
 \usepackage{amsmath,amsfonts,amssymb}
 \usepackage{pifont}
 \usepackage{rotating}
 \usepackage{graphicx}    
 \usepackage{epstopdf}
 \usepackage{color}
 \usepackage{cancel}
 \usepackage[caption=false]{subfig}
 \usepackage{soul}
 
 \usepackage{hyperref}
 \definecolor{dark-red}{rgb}{0.9,0.15,0.15}
 \definecolor{dark-blue}{rgb}{0.15,0.15,0.4}
 \definecolor{medium-blue}{rgb}{0,0,0.5}
 \hypersetup{
 	colorlinks, linkcolor={black},
 	citecolor={dark-blue}, urlcolor={medium-blue}
 }

\begin{document} 

\title{Coexistence of  magnetic and dielectric  glassy states in alternating kagome and triangular lattice LuBaCo$_{4}$O$_{7}$ cobaltite} 

\author{C.Dhanasekhar}
\email{dsekhar21iitb@gmail.com}
\affiliation{Department of Physics, Indian Institute of Technology Bombay, Mumbai 400076, India}
\affiliation{Department of Physics, National Sun Yat-sen University, Kaohsiung 80424, Taiwan}
\affiliation{Center of Crystal Research, National Sun Yat-sen University, Kaohsiung, 80424, Taiwan}
\author{D. Chandrasekhar Kakarla}
\affiliation{Department of Physics, National Sun Yat-sen University, Kaohsiung 80424, Taiwan}
\author{Archana kumari}
\affiliation{Department of Physics, Indian Institute of Science, Bengaluru 560012, India} 
\author{Monika Jawale}
\affiliation{Department of Physics, Indian Institute of Technology Bombay, Mumbai 400076, India} \author{Ronit Hindoddikar}
\affiliation{Cryogenic Enginerring Centre, Indian Institute of Technology, Kharagpur -721302, India}
\author{A. Tiwari}
\affiliation{Department of Physics, National Sun Yat-sen University, Kaohsiung 80424, Taiwan}
\author{Patri Tirupathi}
\affiliation{University Department of Physics, Tilk Manjhi Bhagalpur University, Bhagalpur 812007, India}
\author{Cang Ting Lai}
\affiliation{Department of Physics, National Sun Yat-sen University, Kaohsiung 80424, Taiwan}
\author{ Mitch M. C. Chou}
\affiliation{Center of Crystal Research, National Sun Yat-sen University, Kaohsiung, 80424, Taiwan}
\author{A. Venimadhav}
\affiliation{Cryogenic Enginerring Centre, Indian Institute of Technology, Kharagpur -721302, India}
\author{H. D. Yang}
\email{yang@mail.nsysu.edu.tw}
\affiliation{Department of Physics, National Sun Yat-sen University, Kaohsiung 80424, Taiwan}
\affiliation{Center of Crystal Research, National Sun Yat-sen University, Kaohsiung, 80424, Taiwan}
\author{Praveen Chaddah}
\email{chaddah.praveen@gmail.com}
\affiliation{UGC-DAE Consortium for Scientific Research, Khandwa Road, Indore 452001, India}
\author{A. V. Mahajan} 
\email{mahajan@phy.iitb.ac.in}
\affiliation{Department of Physics, Indian Institute of Technology Bombay, Mumbai 400076, India}


\begin{abstract}
To date, the alternating kagome and triangular lattice cobaltites, i.e., RBaCo$_{4}$O$_{7}$ (R = Ca, Y, and rare earth), have been well studied due to their large structural distortions, anisotropic exchange interactions, chiral spin liquid states, and giant multiferroic properties. Here, we report the co-existence of magnetic and dielectric glassy states in LuBaCo$_{4}$O$_{7}$ cobaltite below 50\,K. AC magnetization studies  show an absence of conventional spin freezing phenomena. \,The cooling and heating in unequal field (CHUF), thermal cycling of magnetization, and time-dependent magnetization studies at the low temperature ($T$) show the presence of the magnetic glassy state. The $T$ dependent dielectric constant ($\epsilon^{'}$) exhibits a strong frequency-independent response at the first-order structural phase transition temperature $T$ $=$ 160\,K; (trigonal $P31c$ to monoclinic $Cc$) and also significant features at the  $T$ $ \approx $ 110\,K (monoclinic $Cc$ to orthorhombic $Pbn2_{1}$) phase transition. Further, $\epsilon^{'}$ shows a frequency-independent peak at 43\,K ($Pbn2_{1}$) and also dipolar glassy features below 20\,K ($Cc$). The non-equilibrium magnetic glassy dynamics and dipolar glassy state at low-$T$ arises from the kinetic arrest of $Cc$ and $Pbn2_{1}$ phases. From the dielectric probe, we are able to clearly distinguish the kinetically  arrested phases at low-$T$, whereas the bulk magnetization studies are unable to do as the arrested phases have low magnetic moments.

\end{abstract}
		
\date{\today}

\maketitle

\section{Introduction}

RBaCo$_{4}$O$_{7}$ (R $=$ Ca, Y, In and rare earth) oxides, known as the R-114 cobaltites family, act  as a model system for studying the interplay between structural, magnetic, thermal and electrical properties\cite{Khomskii,Kocsis2024,Chris,Sarkar2010,raveau2012,Raveau2010,Avci,kazei2011,Caignaert}.The trigonal structural framework of R-114 consists of alternating two-dimensional kagome  and triangular layers, with cobalt ions sitting at the centers of corner-sharing oxygen tetrahedra. In R-114 oxides, the cobalt ions are in  mixed Co$^{2+}$ and Co$^{3+}$ states to maintain a charge balance and distributed among the two independent sites, located in the kagome and triangular layers. For trivalent R ions (such as In, Y, or a rare earth element) in R-114 oxides, the average Co oxidation state is expected to be 2.25+, with a Co$^{2+}$/Co$^{3+}$ ratio of 3:1 \cite{Hollmann}. The R-114 cobaltites undergo a first-order structural phase transition from trigonal $P31c$ to orthorhombic $Pbn2_{1}$, where the transition temperature ($T$) is denoted as $T_{S1}$. For large R ions, $T_{S1}$ occurs slightly above room-$T$ and decreases to about 100\,K with decreasing size of the R ion \cite{Khomskii,Sarkar2010,Mitchell2009,Chris}. 

At low-$T$ ($T$ $<$ 100\,K), the R-114 cobaltites display diverse magnetic properties, and the magnetic ground state is significantly influenced by the R cation \cite{Sarkar2010}. For example, CaBaCo$_{4}$O$_{7}$ (CBCO) exhibits clear ferrimagnetic order below 70\,K \,\cite{Singh,CDS2017,CDS2016}, InBaCo$_{4}$O$_{7}$ (IBCO) exhibits spin-glass-like freezing \cite{RUAN20151}, spin glass along with short-range  magnetic correlations and also Griffiths phase in DyBaCo$_{4}$O$_{7_{\pm\delta}}$\cite{Kumar,Dhanasekhar_2019} spin liquid-like properties along with long-range antiferromagnetic (AFM) ordering\,($T_{N}$)\,in YBaCo$_{4}$O$_{7}$ (YBCO) \cite{Manuel2009,Khalyavin}. The doping of Fe at Co sites in YBCO leads to the chiral spin liquid ground state \cite{Schweika}. TbBaCo$_{4}$O$_{7}$ shows coexisting long-range and short-range magnetic phases and converted to long-range under an applied field \cite{bludov2009,szymczak2010}. Extensive doping studies have been carried out in CBCO at Ca and Co sites, which suppress the long-range ferrimagnetic order and induce the cluster or spin glass phases \cite{Seikh2012,CD2020}. 

In previous studies, it was reported that LuBaCo$_{4}$O$_{7}$ (LBCO) shows  two successive first-order phase transitions, i.e., at $T$ $=$ 160\,K; from the trigonal $P31c$ phase to the monoclinic $Cc$ phase and at $T$ $ \approx $ 110\,K from the monoclinic $Cc$ phase to the orthorhombic $Pbn2_{1}$ phase \cite{Avci,Soda2007}. The $P31c$ to $Cc$ phase transition was clearly evident from electrical resistivity and specific heat measurements. Conversely, the $Cc$ to $Pbn2_{1}$ phase transition features were shown in specific heat and magnetization ($M$) studies, and at this transition, $M$ shows a large thermal hysteresis. Among R-114 cobaltites, the monoclinic $Cc$ phase is unique to LBCO and is not observed/reported in other materials of this family. At low-$T$, $M$ studies observed a significant peak at 48\,K, which is not related to any long-range magnetic order but is assigned to the representation of the metastable supercooled phase. Extensive synchrotron and neutron powder diffraction studies at low-$T$ reported the  coexistence of $Cc$ and $Pbn2_{1}$ phases and their phase fraction is strongly dependent on the cooling rate \cite{Avci}. Slow cooling rates ($\frac{dT}{dt}$ $<$ 3 K/min) resulted in coexisting $Pbn2_{1}$ and $Cc$ phases, with varying fractions dependent on the cooling rate. Fast cooling ($\frac{dT}{dt}$ $\geq$ 3.5 K/min) suppressed the low-$T$,  $Pbn2_{1}$ phase and allowed the high-$T$, $P31c$ phase to convert to a metastable $Cc$ phase below 160 K, which is expected to be  stable down to 2 K. Further, neutron powder diffraction studies showed absence of long-range magnetic ordering in Lu-114 down to 2\,K, and additionally showed broad diffuse magnetic scattering intensities at low-$T$, indicating short-range magnetic correlations at low-$T$. However, single crystal neutron diffraction studies on LBCO at low-$T$ showed a commensurate and incommensurate-like modulation of the cobalt moments on the kagome and triangular lattices \cite{Soda2007,Soda2021}.

Although neutron studies in LBCO revealed short-range magnetic correlations and coomensurate or incommensurate-like phases at low$T$, bulk $M$- measurements showed a significant peak at 48\,K and broad thermal hysteresis between 70\,K and 110\,K. This observation motivated us to conduct detailed $M$ studies to investigate the presence of glass-like features in LBCO at low-$T$. If there are glassy features, the nature of the glassy state, and whether the glassy state is associated with first-order phase transitions and is influenced by the different phases. To answer these questions, we have examined the magnetic ground state of LBCO using various magnetic measurement protocols. Our results show that the first-order structural phase transitions in LBCO are kinetically arrested at low-$T$ and form a nonequilibrium magnetic state, i.e., a magnetic glass. Furthermore, due to the noncentrosymmetric crystal structure and significant structural distortions, the R-114 cobaltites are anticipated to show strong correlations among their structural, magnetic, and dielectric properties \cite{Raveau2010,Caignaert,Kocsis2024,Sarkar2010}. In this family of materials, it is reported that CBCO cobaltite shows large structural distortions that stabilize the polar orthorhombic $Pbn2_{1}$ phase up to 4\,K and these distortions stabilize the long-range ferrimagnetic ground state (FiM) up to 70\,K. The large structural distortions, together with long-range FiM order, further drive the giant spin-order-driven electric polarization in CBCO below 70\,K \cite{Raveau2010,Caignaert}. In this study, for the first time, we demonstrate the strong correlations among these properties, even in the absence of the large structural distortions and long-range magnetic ground state. Detailed dielectric measurements demonstrate structural phase transistions strongly coupled with dieleltric properties and show a dipolar glassy state at low-$T$, which arises from the supercooled and kinetically arrested metastable phase, that is, monoclinic $Cc$.

\section {Experimental Details}

The LuBaCo$_{4}$O$_{7}$ (LBCO) sample was prepared by a solid-state reaction method. The starting materials of Lu$_{2}$O$_{3}$ (99.995\%, Alfa Aesar), BaCO$_{3}$ (99.9\%, Alfa Aesar) and Co$_{3}$O$_{4}$(99.7\%, Alfa Aesar) were mixed in the appropriate ratio. The homogeneous powder was placed in a platinum crucible with a lid and heated in air at 900~$^0$C for 12 h and at 1150~$^0$C for 48\,h with intermediate grindings. To stabilize the 114 phase, the sample was directly quenched in air from 1150 ~$^0$C to room temperature, which is necessary, similar to the other cobalt compounds in this family \cite{Seikh2012,CD2020,CDS2016}. The room-temperature powder X-ray diffraction (XRD) on LBCO samples was performed using a PANalytical X$^{'}$pert PRO diffractometer equipped with Cu-$K\alpha$ radiation radiation ($\lambda$ $=$ 1.54182\, $\mathring{A}$). The experimental XRD data is analyzed through the Rietveld refinement method using FullProf Suite software. All the DC and AC magnetic measurements were performed using a Magnetic Property Measurement System (MPMS, SQUID VSM), Quantum Design, USA. All the DC magnetization measurements were performed during both cooling and heating at rates of 3 K/min, and the AC magnetization measurements were performed under heating rates of 1 K/min. Temperature-dependent dielectric ($\epsilon^{'}(T)$) measurements were performed using an LCR meter (Agilent E4980A). All the dielectric and pyroelectric current measurements were conducted using the Janis cryogenic and Quantum Design MPMS systems supplemented by a homemade multifunctional probe attached to the MPMS. The pyroelectric current ($I_{p}$) was measured using a Keithley 6517A electrometer under conventional electric poling and bias poling electric field methods during warming temperature cycles \cite{Ajay2022,CDS2017,Pal}. A pellet of area 48.9 mm$^2$ and thickness of 0.3 mm was used for the dielectric and pyroelectric current measurements. The $\epsilon^{'}(T)$ measurements are performed under an ac excitation field of 1 V with silver paint applied to both sides of the sample to act as the electrodes. The cooling and heating-dependent $\epsilon^{'}(T)$ measurements were performed under 2 K/min and 1 K/min rates, and these temperature ramping conditions are specified in the relevant main text.Resistivity measurements were performed using standard four-probe geometry in a Janis cryogenic PPMS system with a Keithley 2400 source meter.

\section {Results and discussion}
\subsection{Room temperature X-ray diffraction studies}
\begin{figure}[h!]
\includegraphics[width=\linewidth]{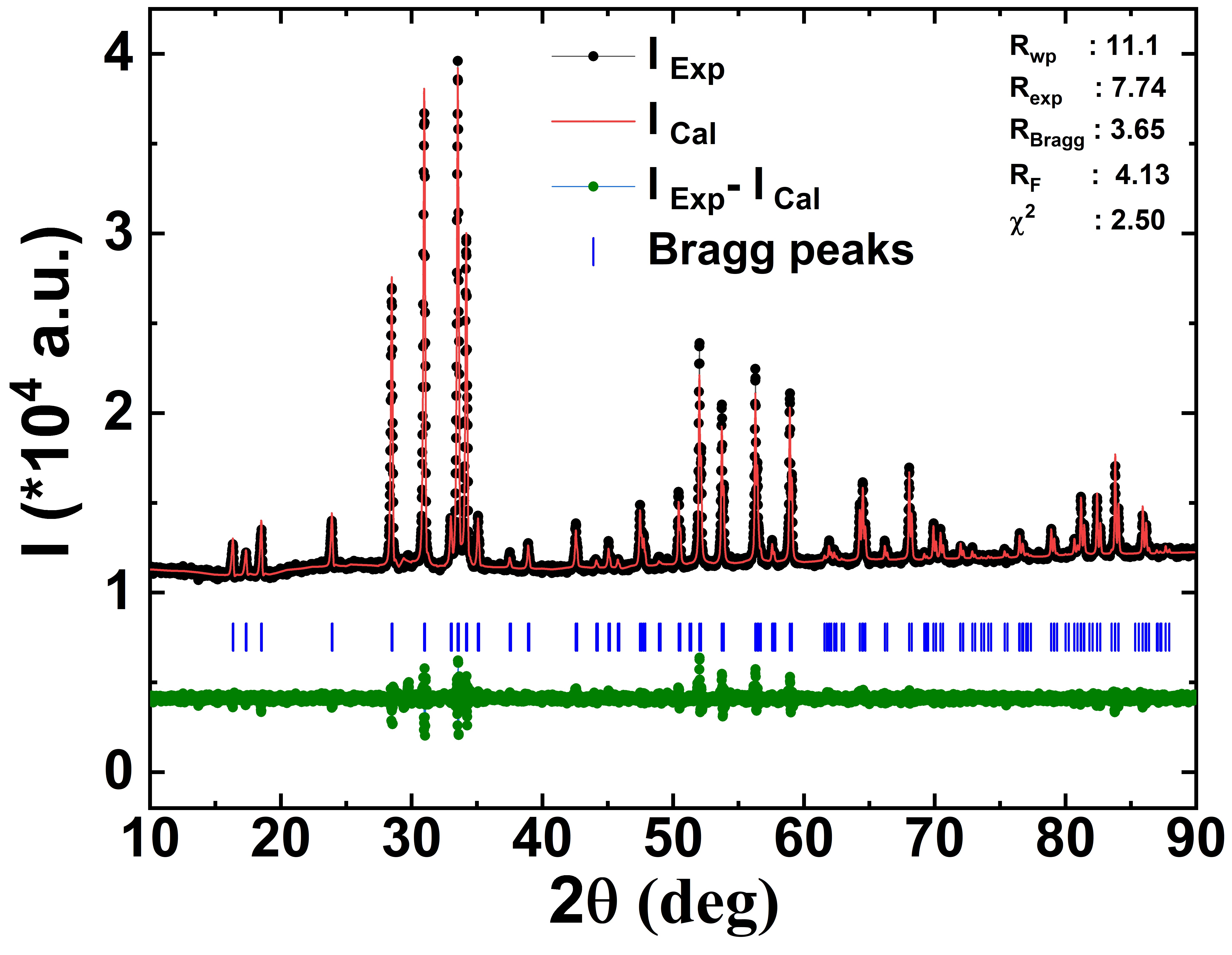}
\centering
\caption{Room temperature powder X-ray diffraction data of the LuBaCo$_{4}$O$_{7}$ sample along with the corresponding Rietveld refinement model.}
\label{Fig1}	
\end{figure}

The room-temperature X-ray diffraction (XRD) patterns of the LuBaCo$_{4}$O$_{7}$ sample, along with Rietveld refinement, are shown in Fig.\,\ref{Fig1}.Rietveld analysis confirms that LBCO belongs to the trigonal $P31c$  space group with the lattice parameters $a$ $=$ $b$ = 6.26350 (03) $\mathring{A}$, $c$ = 10.22467 (10) $\mathring{A}$, and $V$ = 347.388 (4)\, $\mathring{A}^{3}$, respectively. The obtained lattice parameters are within the range of reported values \cite{Avci} and the refinement agreement factors are given in the inset of Fig.\,\ref{Fig1}. 

\subsection{DC magnetization  studies}

\begin{figure}[h!]
\includegraphics[width=0.95\linewidth]{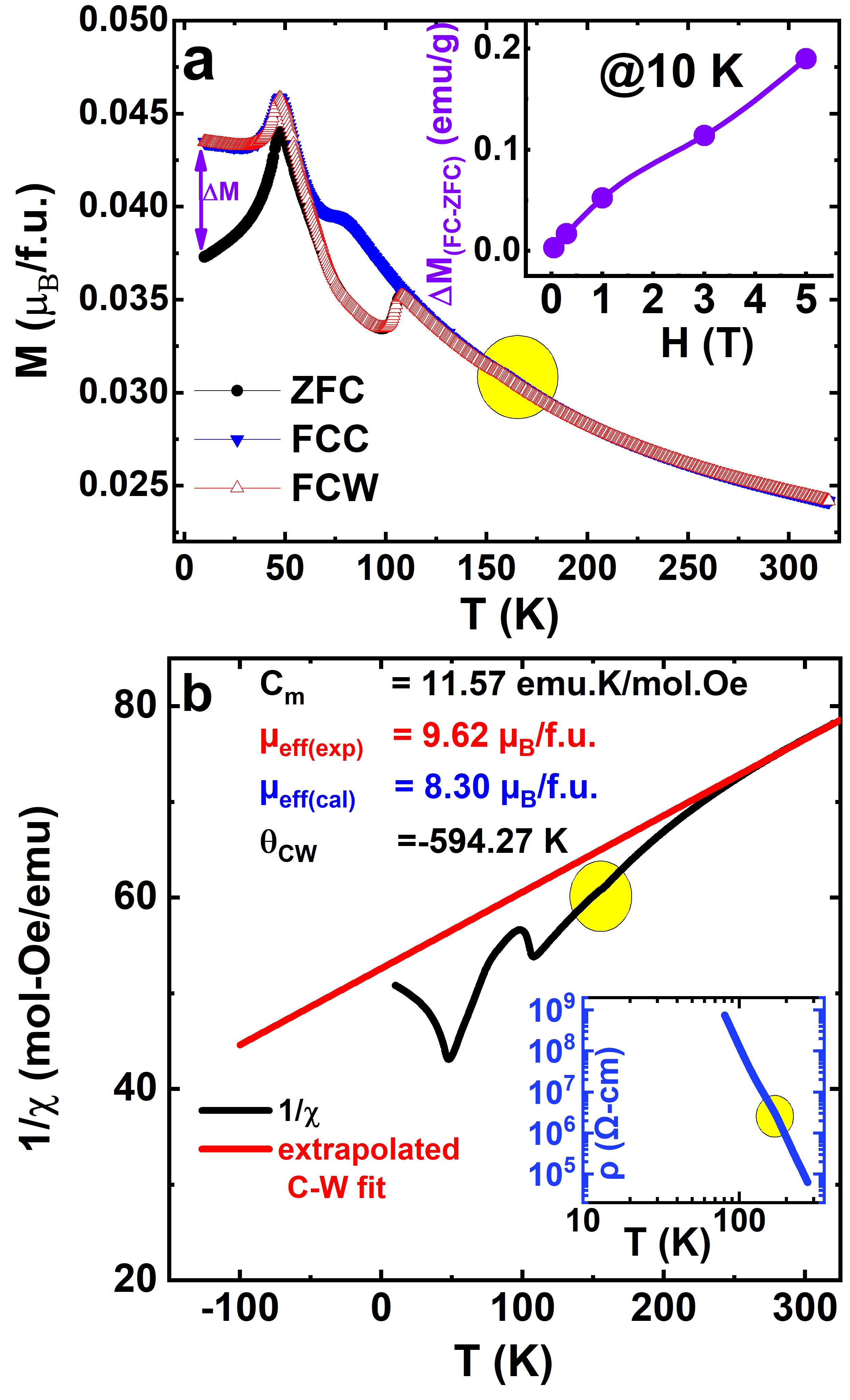}
\centering
\caption {(a) shows the  $M$ vs. $T$ curves of the LuBaCo$_{4}$O$_{7}$ sample measured using ZFC, FCC, and FCW protocols under $H$ = 1\,T. (b) shows the 1/$\chi$ vs. $T$ data along with the extrapolated C-W fitting.\,The inset of (a) shows the $\Delta\,M$ vs.$H$ measured at 10\,K and inset of (b) shows the $\rho$ vs.$T$ in log-log scale}.
\label{Fig2.jpg}	
\end{figure}

The $T$-dependent magnetization ($M$) of LBCO was measured under a magnetic field ($H$) of 1 T under Zero-Field-Cooled (ZFC), Field-Cooled Cooling (FCC), and Field-Cooled Warming (FCW) protocols, as shown in Fig.\ref{Fig2.jpg}(a). The $M$\,vs.\,$T$ curves curves show several anomalies with  decreasing $T$. All anomalies  match well with previous polycrystalline and single crystal studies of LBCO \cite{Avci,Soda2007}. With decreasing $T$, the $M$ vs. $T$ curves show a small kink at 160\,K (yellow circle in Fig.\ref{Fig2.jpg}(a and b), and this transition was not observed in previous $M$ studies. The $\rho$ vs.\,$T$ measurements show a slope change at $T$ $=$ 160\,K (inset of Fig.\ref{Fig2.jpg}(b)), similar to earlier resistivity measurements\,\cite{Avci}. However, specific heat and neutron scattering studies showed changes at this $T$, owing to the structural transition from trigonal P\,31c to monoclinic Cc. Further, as $T$ decreases, the $M$ vs. $T$ (ZFC/FCW (heating) and FCC (cooling) protocols) show a large thermal hysteresis in the range of 60 - 110\,K ($\sim $ 20\,K). At low-$T$, the $M$ vs. $T$ curves show a clear peak at 48\,K and strong irreversibility between the ZFC and FCC/FCW  curves below 48\,K. The difference between the ZFC and FCC magnetization, i.e., $\Delta\,M$ ($M_{FC}$\,$-$\,$M_{ZFC}$) at 10\,K increases with $H$ and is shown in the inset of Fig.\ref*{Fig2.jpg}(a). The high-$T$ magnetic susceptibility data was fitted with the Curie-Weiss (CW) law 

\begin{equation}
\chi = \frac{C_m}{T - \theta_{\text{CW}}}
\end{equation}

where ${C_m}$ denotes the molar Curie constant and $\theta\textsubscript{CW}$ denotes the C-W temperature. The fitting gives a $\theta_{\text{CW}} = -594.27\,\text{K}$ and $\mu_{\text{eff}} = 9.62 \, \mu_{\text{B}}/\text{f.u.}$. The fitting is shown in Fig.\ref{Fig2.jpg}(b), which clearly deviates below 220\,K, Such behavior was also observed in other compounds in this family \cite{ChaponPRB,Seikh2012,DEY}, where the deviation is caused by the short-range magnetic correlations associated with the kagome and triangular lattice arrangements of Co ions together with the structural phase transistions at high-$T$. The high spin state of $Co^{2+}$ ($s$ = $\frac {3} {2}$) and $Co^{3+}$ ($s$= $2$) ions in an tetrahedral crystal field gives an effective spin-only moment as $\mu_{\text{cal}} = \sqrt{3\mu^{2}_{\text{Co}^{2+}} + \mu^{2}_{\text{Co}^{3+}}} = 8.30 \, \frac{\mu_{B}}{\text{f.u.}}$, where $\mu_{Co^{2+}}$\,=\,3.87 $\mu_{B}$ and $\mu_{Co^{3+}}$\,=\,4.90 $\mu_{B}$, respectively. The calculated moments from the C-W fitting are higher than the spin values, suggesting that the orbital moments of $Co^{3+}$ and $Co^{2+}$ ions may not be quenched.

\subsection{AC magnetization  studies}

\begin{figure*}
\includegraphics[width=0.8\linewidth]{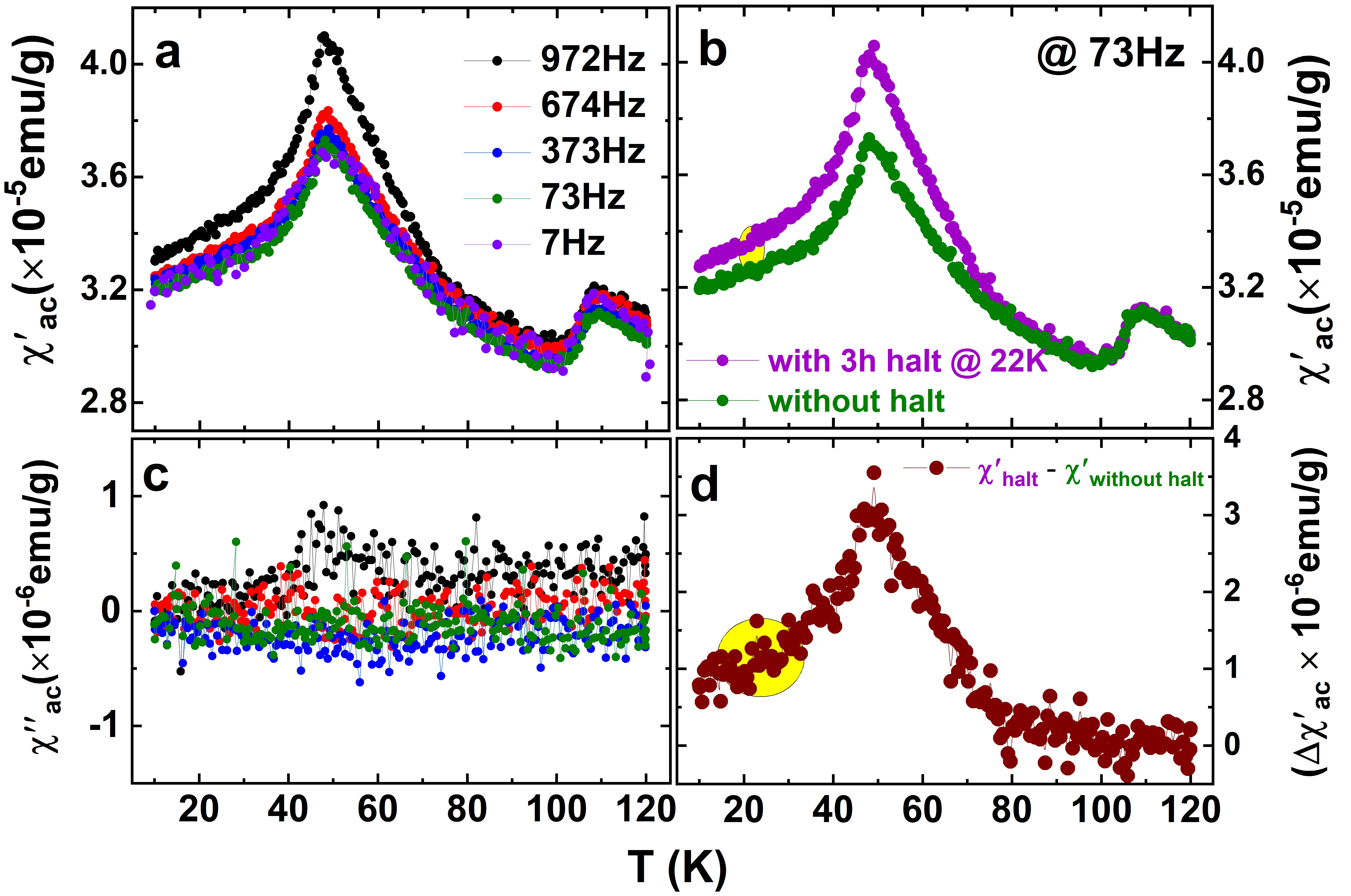}
\centering
\caption {(a) and (c) show the temperature- and frequency-dependent real ($\chi^{'}$) and imaginary  ($\chi^{''}$) components of AC magnetization of the LuBaCo$_{4}$O$_{7}$ sample. (b) and (d) show the $\chi^{'}$ low-frequency susceptibility data obtained with the standard experimental approach for observing the memory effect.}
\label{Fig3.jpg}	
\end{figure*}

Fig.\ref{Fig3.jpg}\,(a and c) shows the real ($\chi^{'}$) and imaginary ($\chi^{''}$) components of AC magnetization, measured for the different frequencies of LBCO in the $T$ interval of 10\,-120\,K. For all frequencies, $\chi^{'}$ shows a sharp and frequency-independent maximum close to 48\,K and $\chi^{''}$ is approximately a straight line. The sharp peak of $\chi^{'}$ at 48\,K matches with the DC $M$ peak observed at this $T$.\, Further, a frequency-independent and step-like peak was observed near 110\,K, where the first-order structural phase transition is observed. Further, we also measured $\chi^{'}(T)$ with (3 hours) and without waiting at $T$\,$=$\,22\,K for the frequency of 73\,Hz and results are shown in Fig.\ref{Fig3.jpg}(b). The $\chi_{t= 3h}^{'}$ and $\chi_{t = 0h}^{'}$ (ref) both were measured while heating the sample with a heating rate of 1 K/min. In these measurements, the main difference is in the cooling of the sample. In case $\chi_{t= 3h}^{'}$, the temperature is stopped at 22\,K for 3\,h (waiting time), whereas in $\chi_{t = 0h}^{'}$ (ref), the $T$ was not stopped at 22\,K. Interestingly, the magnitude of $\chi_{t= 3h}^{'}$ is greater than $\chi_{t = 0h}^{'}$ (ref), and both the curves overlap above 70\,K (Fig.\ref{Fig3.jpg}(b)). In conventional spin glasses, the difference in $\triangle$$\chi^{'}$ ($\chi_{t= 3h}^{'}$ $-$ $\chi_{t = 0h}^{'}$ (ref)) would be expected to shows dip-like features at the waiting temperature \cite{Nordblad,Krishna,Kundu_2020}. The $\triangle$$\chi^{'}$ does not show any such features at 22\,K, instead, it shows a clear peak at 48\,K. The absence of the $\triangle$$\chi^{'}$ dips at the waiting temperature has also been reported in layered EuBaCo$_{2}$O$_{5+\delta}$ ($\delta=0.47$) cobaltite \cite{Kumari_2020}, where conventional spin freezing features also absent. Such, $\chi^{'}$,\,$\chi^{''}$ and $\triangle$ $\chi^{'}$ features rule out the conventional spin glass behavior in the title compound.

\subsection{Kinetic arrest and metastability}

The above AC magnetization results show the absence of the conventional spin-glass behavior and the DC magnetization studies reveal that various anomalies are strongly coupled with structural changes, the FCC and FCW curves show a strong thermal hysteresis. Previous studies predicted that the peak at 48\,K is associated with the metastable supercooled phase \cite{Avci}. However, a detailed understanding of this peak and the nature of the supercooled phase is missing. Here, we used different magnetic measurement protocols, including cooling and heating in unequal fields (CHUF) \cite{Banerjee_2008,Banerjee_2009}, time-dependent magnetization and field-cooled isothermal magnetization studies to shed light on the low-$T$ magnetic phase. CHUF measurements have been extensively used to unravel the phase coexistence through the response of materials to external perturbations \cite{Miao,Roy2009,Roy2006,Roy2014,Chaddah2005,wu2006}.

\begin{figure} [h!]
\includegraphics[width=\linewidth]{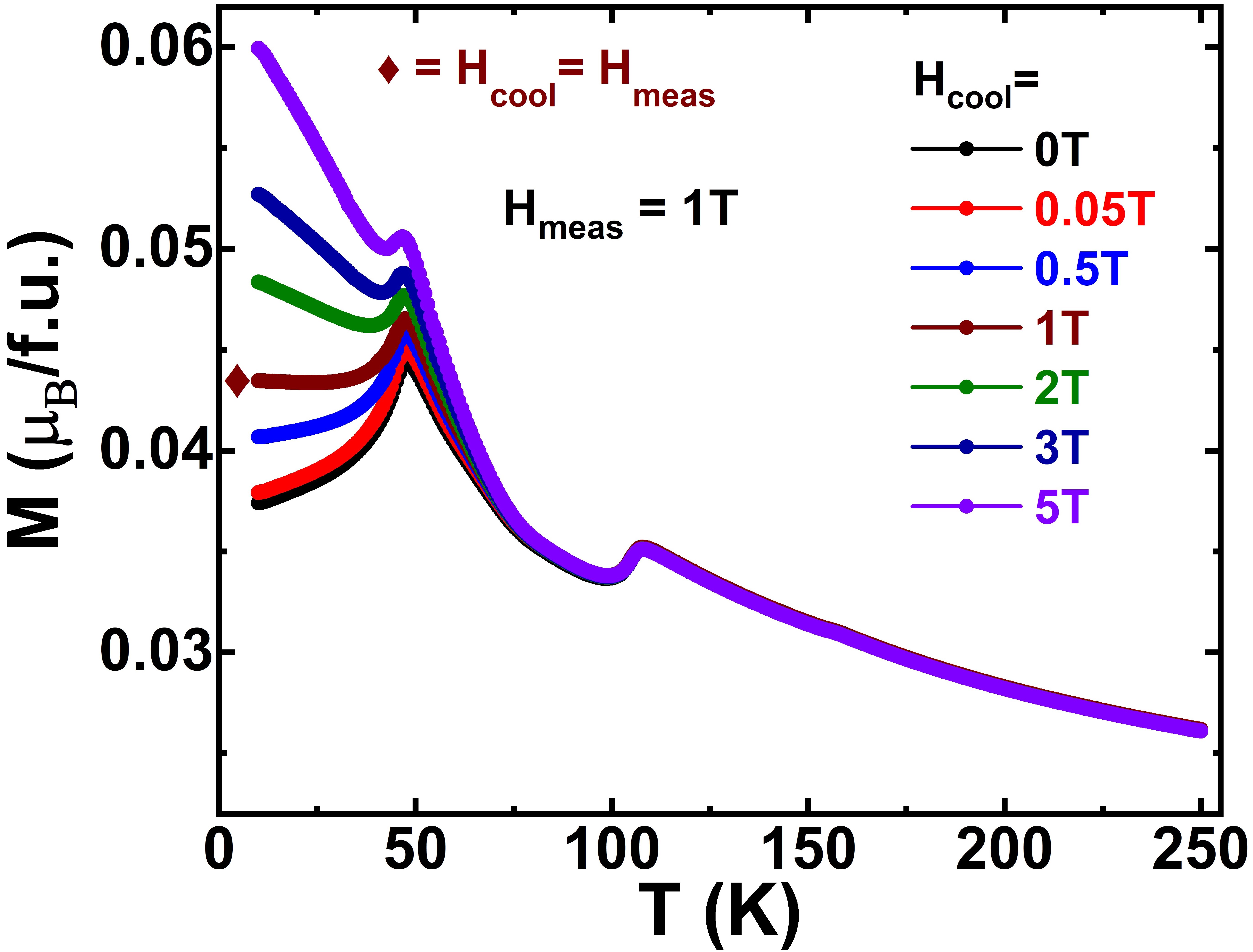}
\centering
\caption {$M\,vs.\,T$ curves of LBCO obtained during heating after cooling under different $H$.}
\label{Fig4.jpg}	
\end{figure}

\begin{figure} [h!]
\includegraphics[width=\linewidth]{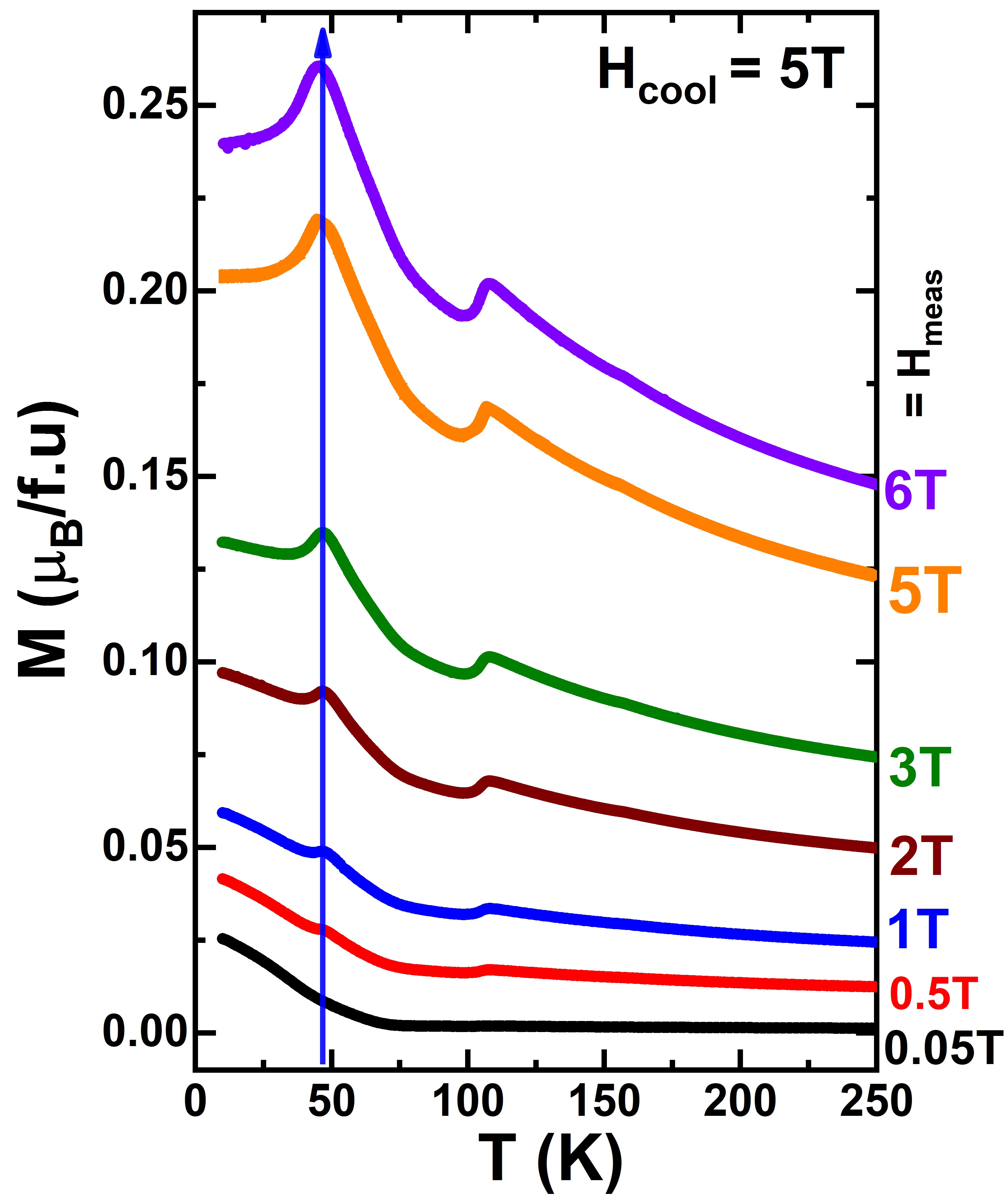}
\centering
\caption {$M\,vs. T$ curves of LBCO obtained during heating after cooling under different $H$.}
\label{Fig5.jpg}	
\end{figure}

Following the CHUF1 protocol (i.e., cooling in unequal fields and heating in same field ), the LBCO sample was cooled from 250 to 10\,K under various cooling magnetic fields ($H_{cool}$), i.e., 0, 0.05, 0.5, 1, 2, 3, and 5 T, respectively. At 10\,K the $H_{cool}$ was changed to the measuring magnetic field ($H_{meas}$) of 1 T and the obtained results are shown in Fig.\ref{Fig4.jpg}. For $T$ $>$ 70\,K, the $M$ behavior of LBCO under  CHUF1  remains the same irrespective of $H_{cool}$. For $T$ $<$ 70\,K, the $M$ curves show a dispersion, which increases with decreasing $T$. The peak position at 48\,K is independent of $H_{cool}$, the behavior of the $M\,vs. T$ curves changes with $H_{cool}$ for $T < 48\,K$. For $H_{cool}$ $<$ $H_{meas}$, $M$ increases from 10\, - 48\,K, for the $H_{cool}$ $=$ $H_{meas}$, the $M$ is almost constant and for the $H_{cool}$ $>$ $H_{meas} $, $M$ decreases. We also used another CHUF protocol, CHUF2 (i.e., cooling in the same field and heating in different fields) to understand the $M$ at low-$T$. In these measurements, the sample was cooled from 200\,K to 10\,K. At 10 K, the $H$ was isothermally reduced to the measuring field $H_{meas}$ and measurement is carried out while heating. We performed these measurements under two different cooling fields, i.e., $H_{cool} $\,=\,1\,T and $H_{cool} $\,=\,5\,T, Fig.(\ref{Fig5.jpg}) shows the $H_{cool} $ = 5\,T. For, $H_{meas=0.05T} $ $<$ $H_{cool}$,low $T$, $M$ decreases continuously and shows a small kink at $T$ = 48\,K. With increasing the $H_{meas}$, the peak at 48\,K evolves more clearly. For, $H_{meas}$ $=$ $H_{cool}$, at low-$T$, $M$ is almost constant and for $H_{meas}$ $>$ $H_{cool} $, at low-$T$, $M$ increases monotonically up to $T$ = 48\,K. Almost similar behavior is observed for the different cooling field, i.e., $H_{cool}$ \,=\,1\,T and in this case, constant $M$, is evident for $H_{meas}$ $=$ $H_{cool}$, i.e., 1\,T.

These CHUF results suggest that the low-$T$ $M$ is strongly dependent on the cooling and measuring fields and is almost constant for $H_{meas}$\,=\,$H_{cool} $. Further, from the CHUF measurements for $H_{cool}$\,$>$\,$H_{meas} $, the $M$ at low-$T$ decreases almost linearly up to 48\,K. This indicates that the frozen or arrested $M$ (while cooling) at low-$T$ is higher and decays fast while measuring in low $H_{meas}$. However, for $H_{cool} $\,$<$\, $H_{meas}$, the $M$ at low-$T$ increases up to 48\,K. This behavior is similar to the ZFC in Fig.\ref{Fig2.jpg}(a), indicating that the $H_{meas}$ is not sufficient to de-arrest the frozen or arrested $M$ (while cooling) at low-$T$. Alternatively, there may be a small or negligible frozen or arrested $M$ (while cooling) at low-$T$. All these observations indicate that the low-$T$ $M$ is metastable and is kinetically arrested while cooling. Further, the kinetically arrested $M$ state reaches a stable magnetization above 70\,K. Here, we note that the de-arrest takes place at 48\,K and the $M$, becomes stable above the 70\,K. 

\begin{figure} 
\includegraphics[width=\linewidth]{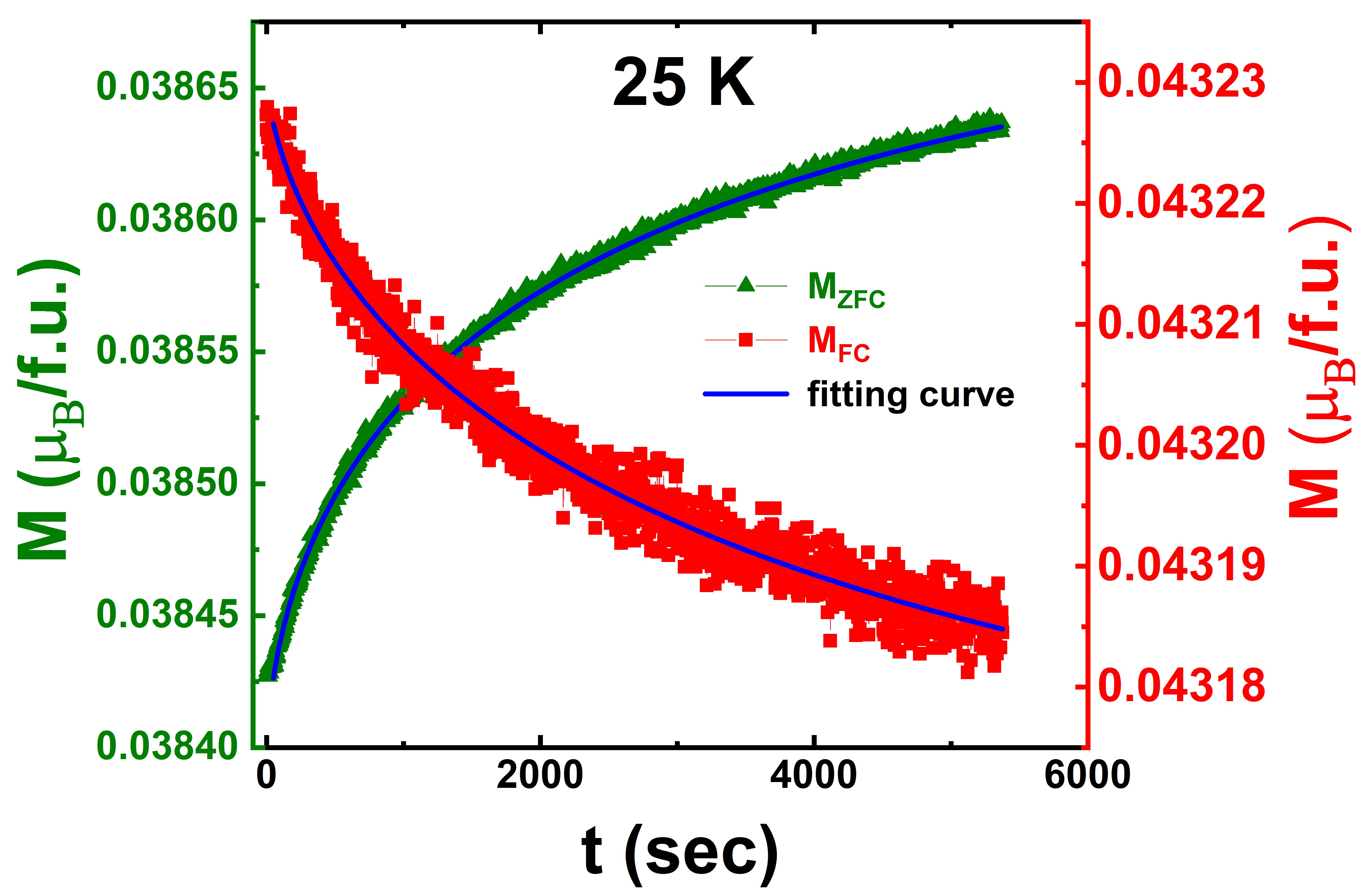}
\centering
\caption {Relaxation of ZFC and FC  magnetization measured at $T$ = 25 K. The solid lines represent the fit using a stretched exponential function in Eq. (1)}
\label{Fig6.jpg}	
\end{figure}

Further, to understand the low-$T$ magnetic state we, performed magnetic relaxation measurements at 25\,K in the ZFC and FC conditions, and the obtained results are shown in Fig.\ref{Fig6.jpg}. In the ZFC measurements, the sample was cooled from 300\,K to 5\,K and at 5\,K, $H$ of 1\,T was applied and the sample was heated to 25\,K. After reaching 25\,K, we waited for 5 minutes for thermal equilibrium and started measuring the $M$ with time ($t$). In the FC measurements, the sample was cooled from 300\,K to 5\,K, under a $H$ of 1\,T and heated from 5\,K to 25\,K. After reaching 25\,K, we waited for 5 minutes for thermal equilibrium and started measuring the $M$ with time. From Fig.\ref{Fig6.jpg}, it can be noted that the $M_{ZFC}(t)$ of LBCO increases from 0.03843 to 0.03864 $\frac{\mu_{B}}{f.u.}$ and the corresponding change in $M$ is 0.54\,$\%$. However, $M_{FC}(t) $ decreases from 0.04323 to 0.04318 $\frac{\mu_{B}}{f.u.}$ and the corresponding change in $M$ is  0.11$\%$. At 25\,K, the initial $M_{FC} $ is 1.1 times of $M_{ZFC} $ indicating that the $M_{FC} $ of LBCO is closer to the equilibrium state, compared to $M_{ZFC} $.

In general, magnetic glassy systems are understood by analyzing their time-dependent magnetization $M(t)$, which can be expressed as \cite{Pallab},

\begin{equation}
M(t) = M_0 \pm M_{g} \exp\left[-\left(\frac{t}{\tau}\right)^{\beta}\right]
\end{equation}

where, $M_{0} $ is the intrinsic $M$, $M_{g}$ is related to a glassy component of $M$,\,$\tau$ is the characteristic relaxation time constant and $\beta$ is the stretching exponent, which has values between 0 ( M(t) is constant, i.e., no relaxation) and 1 ( M(t) relaxes with a single time constant). In this aspect, the value of $\beta$ conveys the dynamics of spins with a very strong to no relaxation limit. Further, the magnitude of $\beta$ relies on the energy barriers, which are involved in the relaxation. Materials, that contain multiple energy barriers, have $\beta$ values between 0 and 1, whereas for a uniform energy barrier, $\beta$\,=\,1. The values of $\beta$ obtained for the LBCO from Eq.(1) to the $M_{ZFC}(t) $ and $M_{FC}(t)$ are 0.55 and 0.62, respectively. The $\beta$ $<$ 1 values in the LBCO signify that the $M$ relaxation evolves through multiple intermediate metastable states.

\begin{figure*}
\includegraphics[width=0.8\linewidth]{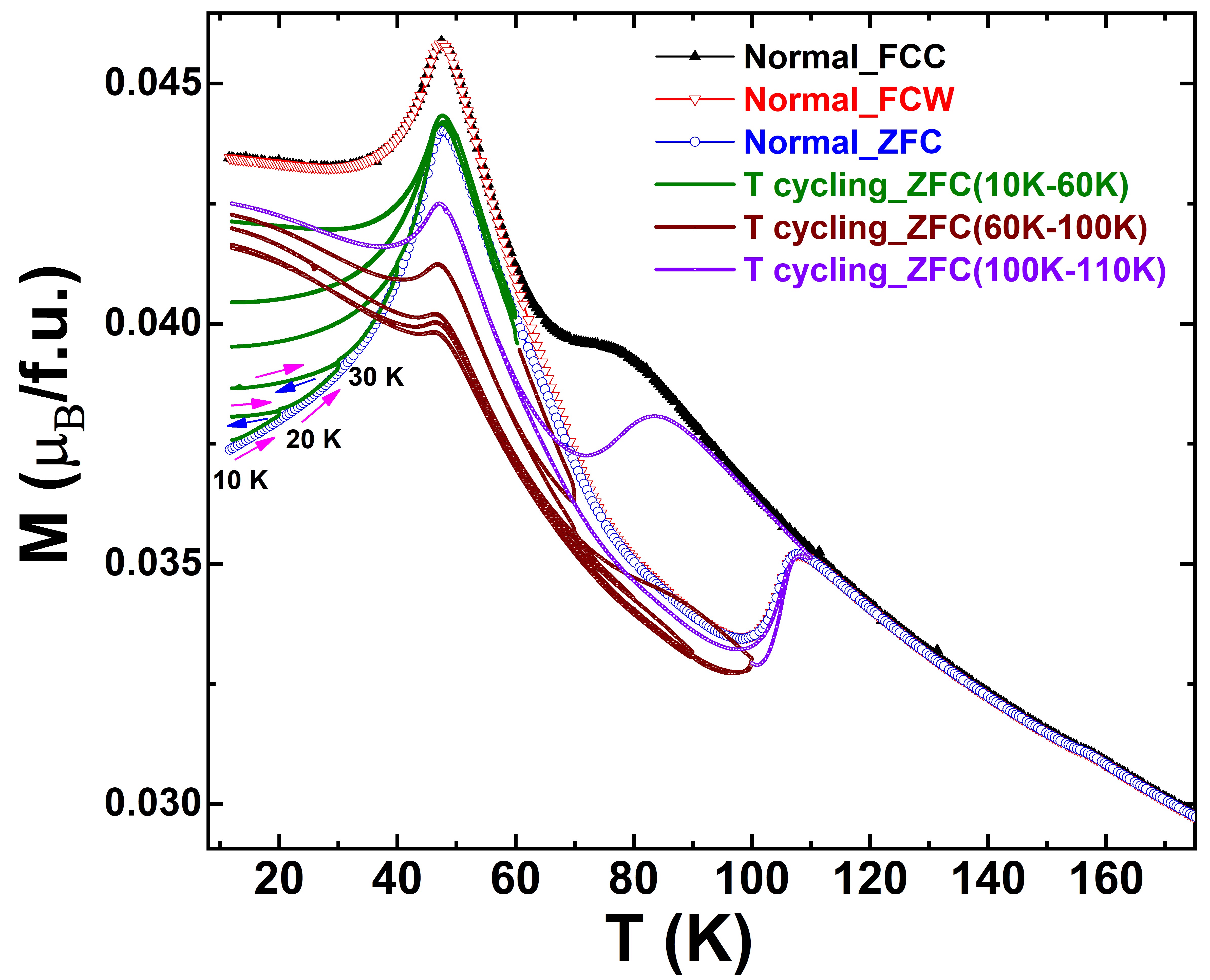}
\centering
\caption {Thermal cycling of $M_{ZFC} \,vs. T$ curves of LBCO are shown along with the normal ZFC, FCC and FCW curves measured under 1\,T. The arrow marks guided the initial thermal cycling process, and the complete thermal cycling was carried out following the same procedure.}
\label{Fig7.jpg}	
\end{figure*}

Fig.\ref{Fig7.jpg} shows the thermal cycling of, $M_{ZFC} \,vs. T$ curves of LBCO measured under 1\,T along with the normal ZFC, FCC, and FCW curves. In the thermal cycling of $M_{ZFC}$ measurements, the sample was cooled from 300\,K to 10\,K and at 10\,K, the $H$ of 1 T was applied. $M$ was measured while increasing and then decreasing the $T$ in steps of 10\,K from 10\,K to 110\,K. For better understanding, the $M$ obatined from the initial thermal cycling is shown in the Fig.\ref{Fig7.jpg} with guided arrow marks between the $T$ intervals of 10\,K-30\,K, and similarly the $M$ was measured up to 110\,K. In these measurements, the $M$ value at low-$T$ increases continuously up to 60\,K, and the thermal cycling between 60\,K and 100\,K causes the $M$ value at low-$T$ to decrease slightly. More interestingly, the magnitude of the peak at 48\,K decreases when thermal cycling is performed between 60\,K to 100\,K, whereas the magnitude of the peak at 48\,K increases when the thermal cycling is performed between 100\,and 110\,K. However, the $M$ at low-$T$ from the thermal cycling of the $M_{ZFC}(T)$ curves does not match with the normal $M_{FCC}(T)$ curves at the highest thermal cycling $T$ of 110\,K. The overall thermal cycling of $M_{ZFC}$ follows the normal $M_{ZFC}$ path and reveals that the thermal hysteresis $T$ region plays a major role in the low-$T$ arrested magnetic phase. These $T$ cycling results are similar to the other magnetic glassy materials, i.e. Gd$_{5}$Ge$_{4}$ and EuBaCo$_{2}$O$_{5+\delta}$ ($\delta=0.47$) cobaltite \cite{Kumari_2020,Roy2006}. These results suggest that, at low-$T$, even at a fixed $H$, thermal cycling can convert some of the supercooled metastable low-$M$ phase ($M_{ZFC}$) to higher-$M$ phase ($M_{FC}$) at equilibrium.

\begin{figure}[h!]
\includegraphics[width=\linewidth]{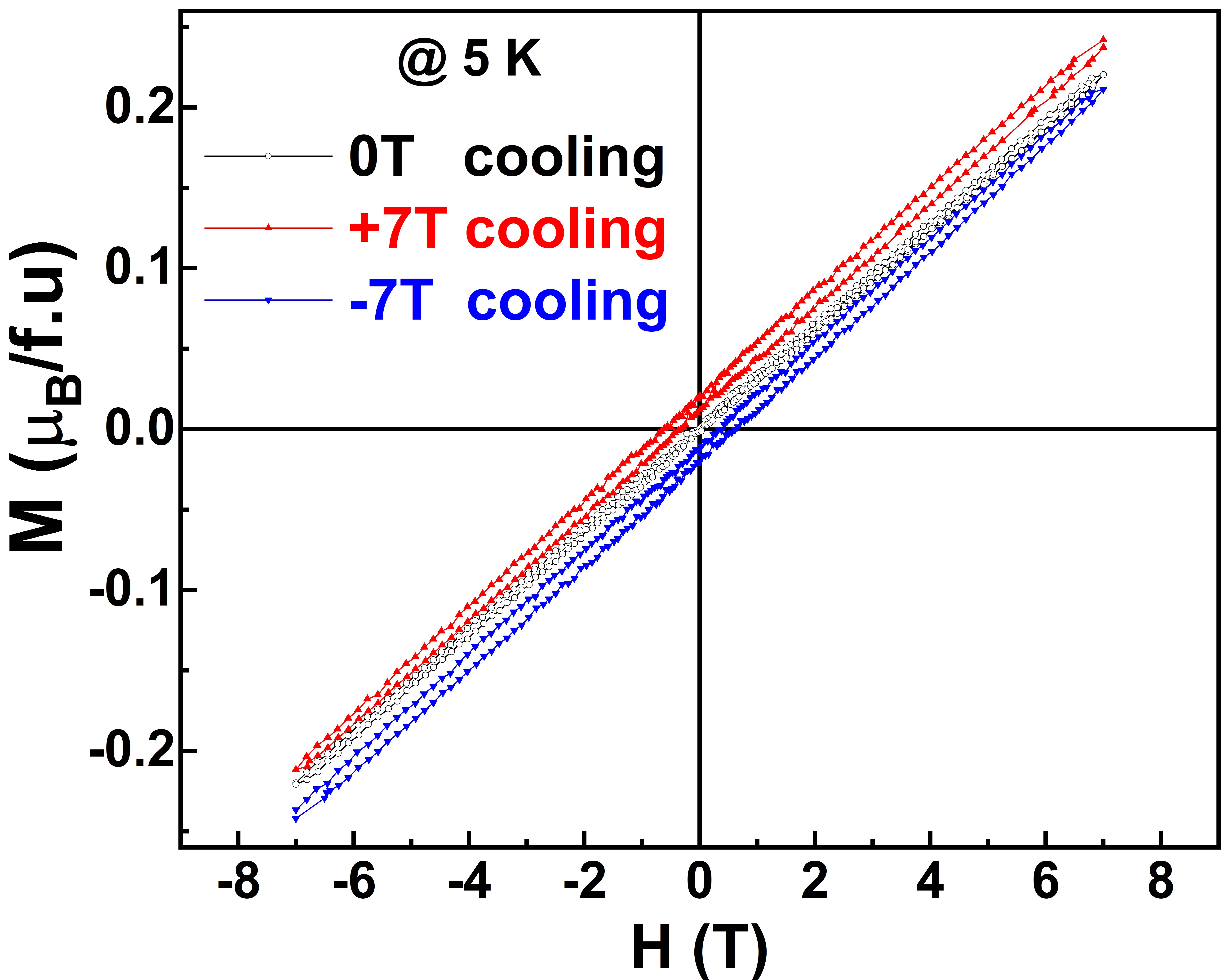}
\centering
\caption {Isothermal ZFC and FC, $M(H)$ loops of LBCO measured at 5\,K.}
\label{Fig8.jpg}	
\end{figure}

The phase coexistence at low-$T$ is further understood from the significant horizontal shift in the $M(H)$ loops measured under $\pm H$ as shown in Fig.\ref{Fig8.jpg}. Such a shift in the $M (H)$ loop under $\pm H$ is known as exchange bias. From the shift of the $M(H)$ loops, the exchange bias field ($H_{E}$= -$[H_{1}+H_{2}/2]$) is obtained using the $H_{1}$ (left  coercive field) and $H_{2}$ (right coercive field). The $H_{E}$ at 7\,T in LBCO is found to be 0.46\,T, which is smaller comapred to the YbBaCo$_{4}$O$_{7}$ (YbBCO). Here, we note that YbBCO shows AFM ordering, whereas LBCO does not \cite{DEY}.

\section {Dielectric studies}

\begin{figure*}
\includegraphics[width=0.9\textwidth]{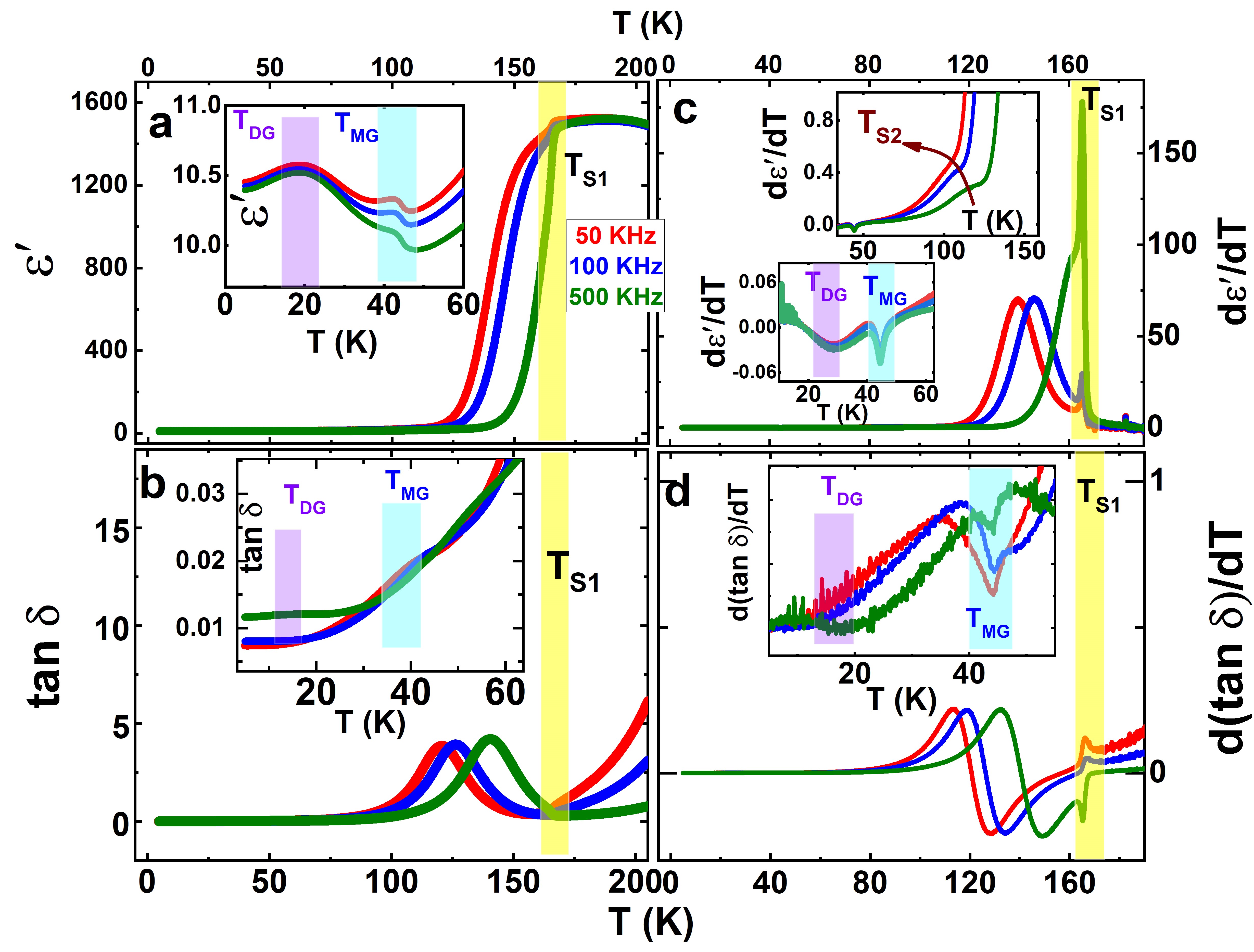}
\centering
\caption {The main panel of (a) and (b) shows the $T$-variation of the $\epsilon^{'}$ and $tan\,\delta$ at different frequencies. The inset of (a) and (b) shows an enlarged view of $\epsilon^{'}$ and $tan\,\delta$ at low-$T$. The main panel of (c) and (d) shows the $\frac{d\epsilon^{'}}{dT}$ and $\frac{d\,tan\,\delta}{dT}$ curves and its insets shows the zoom view at low temperatures. The various transitions are highlighted with different colors and discussed in detail in the main text.}
\label{Fig9.jpg}	
\end{figure*}

The Fig. \ref{Fig9.jpg} (a) and (b) show the $T$ variation of the real part of the dielectric constant ($\epsilon^{'}$) and the corresponding loss tangent ($tan\,\delta$) of the LBCO measured under different frequencies ($f$) with heating rate of 2\,K/min. In the $T$ range of 110\,-\,180\,K the $\epsilon^{'}$ shows a step-like increase and such behavior has also been reported in YBaCo$_{4}$O$_{7}$ (YBCO) and DyBaCo$_{4}$O$_{7}$ (DBCO) cobaltites of the same family \cite{Dhanasekhar_2019}. The stepwise increase in $\epsilon^{'}$ is accompanied by a strong relaxation peak in $tan\,\delta$ (Fig. \ref{Fig9.jpg}(b)), indicating thermally activated relaxation behavior. At $T$ = 165\,K,\, $\epsilon^{'}$ shows a clear sharp kink and the corresponding $tan\,\delta$ also shows a change in slope, at this $T$. The change at $T$ = 165\,K is clearly visible from the first derivative of the $\epsilon^{'}$ and  $tan\,\delta$ with  $T$ as shown in Fig. \ref{Fig9.jpg} (c) and (d) with $T_{S1}$. The $T_{S1}$, closely matches the $M$ changes observed in $M$ curves at $T$ = 160\,K (Fig.\ref{Fig2.jpg}(a) and (b)) and in earlier specific heat and resistivity  \cite{Avci}. Here, we assigned the $T_{S1}$ to the structural transition from trigonal $P31c$ to the monoclinic $Cc$ phase. The step-like increase and strong dispersive nature of $\epsilon^{'}$ in the $T$ interval of 100\,K to 160\,K masks the features of the monoclinic $Cc$ to orthorhombic $Pbn2_{1}$ phase transition at 110\,K. Hence, at this $T$, the $\frac{d\epsilon^{'}}{dT}$ curves show a slight dispersive nature\,(\,upper inset of Fig. \ref{Fig9.jpg}(c)).

For $T$ $<$ 100\,K, $\epsilon^{'}$ decreases up to $T$ $=$ 55\,K and starts rising below 55\,K, which is shown in the inset (zoom view of low-$T$) in Fig.\ref{Fig9.jpg}(a). At $T$ = 43 K, $\epsilon^{'}$ shows a frequency-independent peak, which is close to the magnetic peak $T$ shown in Fig.\ref{Fig2.jpg}(a). With a further decrease in $T$, the $\epsilon^{'}$ shows a broad hump centered at 20\,K, where the hump is dispersive with frequency; $tan\,\delta$ shows frequency-independent and dependent features below 55\,K, which are shown in the inset of Fig. \ref{Fig9.jpg}(b). The frequency-independent features at $T$ = 43 K are clearly shown in the inset of Fig. \ref{Fig9.jpg}(c) and (d) through the  first $T$ derivative of $\epsilon^{'}$ and $tan\,\delta$. The low $T$ frequency-dependent features of LBCO are discussed later. At low-$T$ ($T$ $<$ 100 K), the magnitude of the $\epsilon^{'}$  in LBCO is in good agreement with the previously reported values of well-studied multiferroic materials in this family, such as CBCO \cite{CDS2017,Singh}. The insulating nature of LBCO at low-$T$ (inset of Fig.\ref{Fig2.jpg}(b) and also see impedance studies Fig.\ref{Fig12.jpg} for details), along with the low $\epsilon^{'}$  and low $tan\,\delta$ values below 60 K, excludes the possibility of Maxwell-Wagner-like relaxation features. However, these relaxation features, appear at high-$T$ , coinciding with the structural phase transitions.

\begin{figure}[h!]
\includegraphics[width=\linewidth]{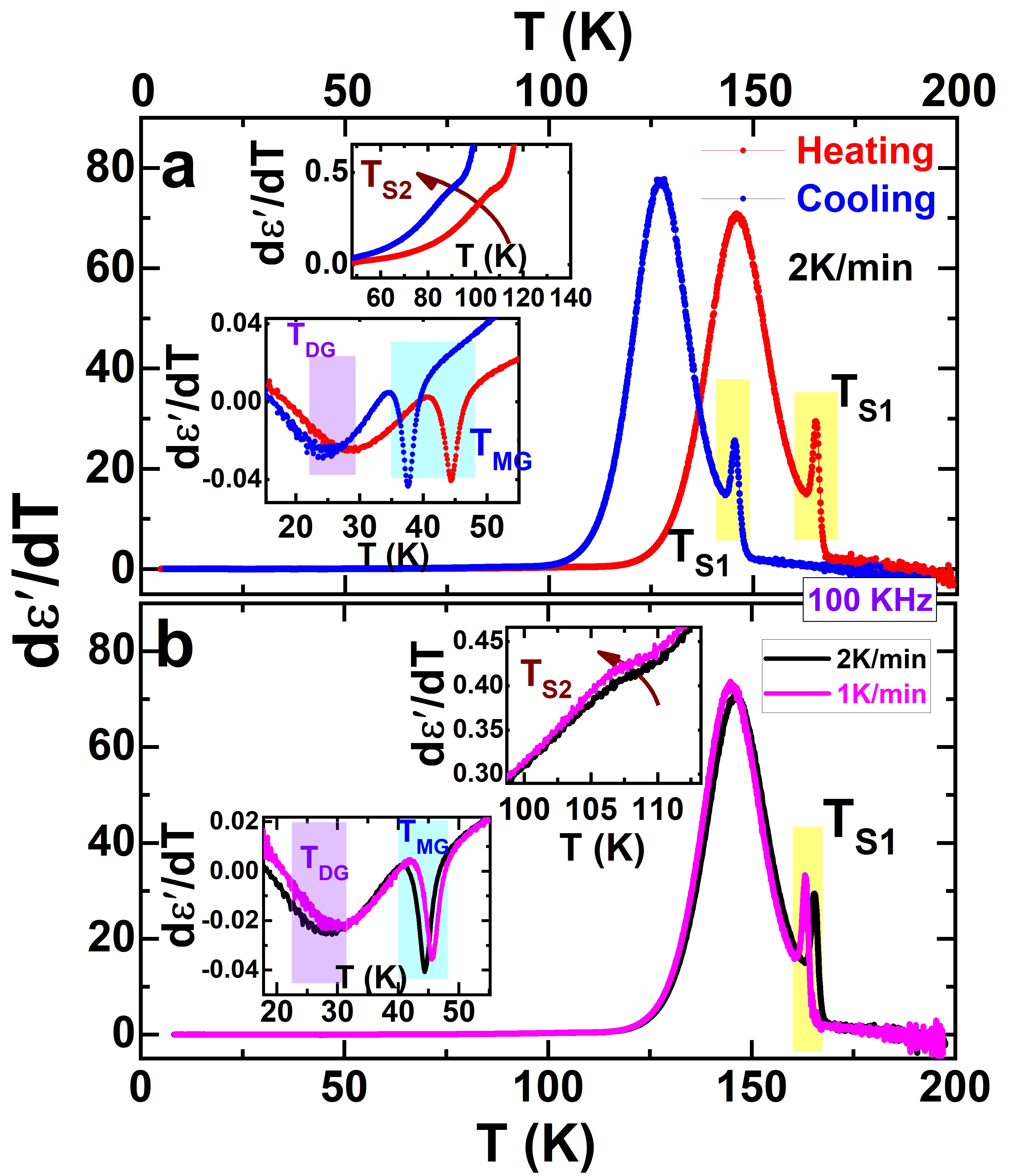}
\centering
\caption {The main panel of (a) shows the $T$ variation of $\frac{d\epsilon^{'}}{dT}$ curves measured under the same heating and cooling rates and (b) shows the $\frac{d\epsilon^{'}}{dT}$ curves measured under different heating rates. The insets show the zoom view of the $\frac{d\epsilon^{'}}{dT}$ curves at low-$T$.}
\label{Fig10.jpg}	
\end{figure}

The $\epsilon^{'}$ vs $T$ curves strongly correlate with structural changes, which are strongly connected to the magnetic ordering $T$. Thus, we also measured the heating and cooling dependence of $\epsilon^{'}$ and $tan\,\delta$ and the corresponding $\frac{d\,\epsilon^{'}}{dT}$ curves are shown in Fig.\ref{Fig10.jpg} (a). Further, the $\epsilon^{'}$ vs $T$ was also measured at different heating rates, 2\,K/min and 1\,K/min, respectively, and the corresponding $\frac{d\,\epsilon^{'}}{dT}$ curves are shown in Fig.\ref{Fig10.jpg} (b). From, Fig.\ref{Fig10.jpg} (a) and (b), it is clear that the $T_{S1}$, $T_{S2}$, $T_{MG}$ and $T_{DG}$ temperatures change significantly with the heating and cooling cycles for the same temperature sweep and also for the different heating rates. This is a strong signature of the first-order structural transition-driven dielectric features.

\begin{figure}
\includegraphics[width=\linewidth]{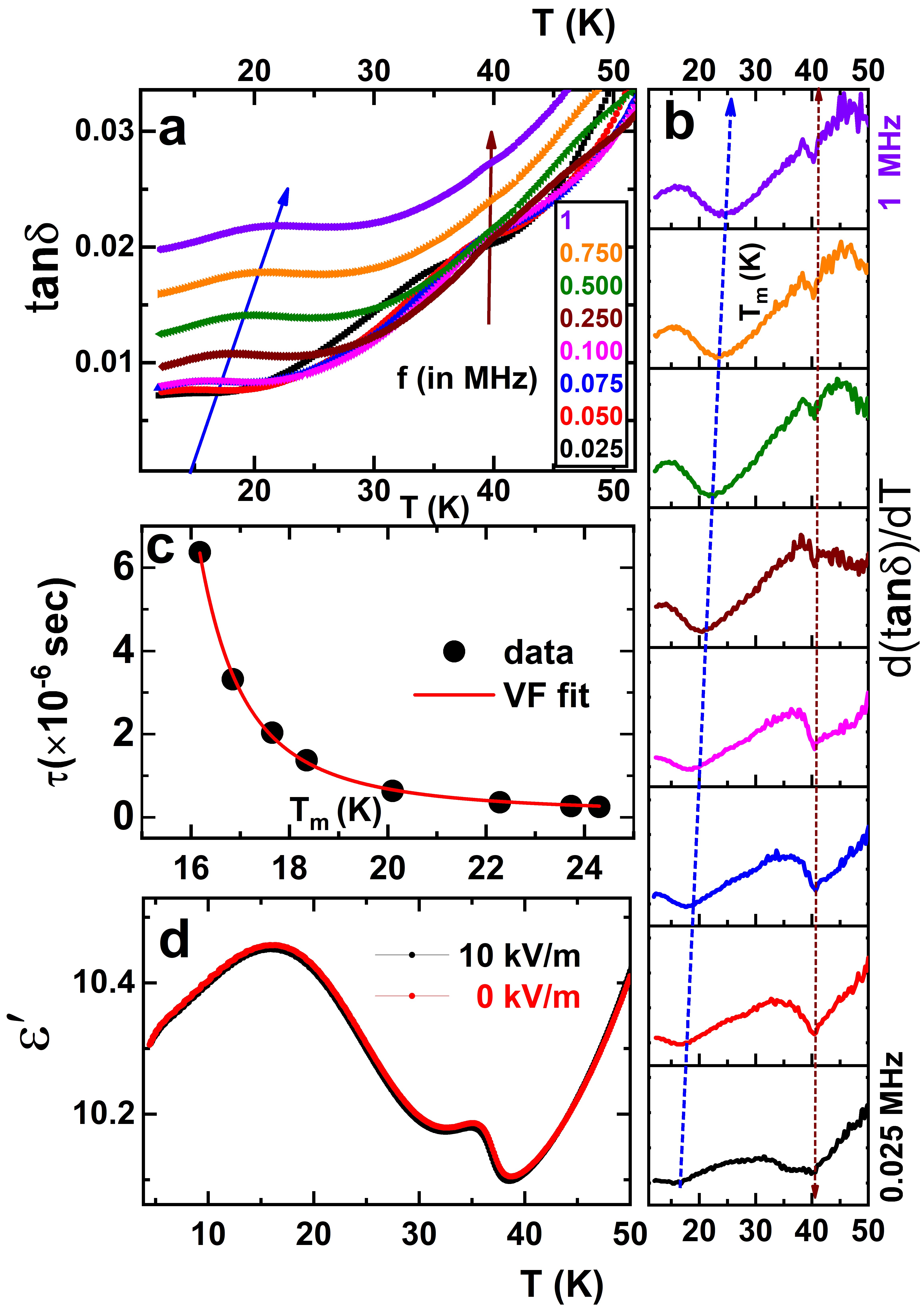}
\centering
\caption {Low-$T$ variation of the $tan\,\delta$ at different frequencies (a), (b) shows the frequency and temperature variation of $T_{m}$ from the $\frac{d\,tan\,\delta}{dT}$ curves. (c) shows the $\tau$ vs. $T_{m}$ plot along with the V-F law fitting, (d) shows the low temperature $\epsilon^{'}$ vs. $T$ curves measured under different DC bias fields.}
\label{Fig11.jpg}	
\end{figure}

Fig.\ref{Fig11.jpg}(a) shows the frequency variation of $tan\,\delta$ at low-$T$. The broad hump at low-$T$ in $tan\,\delta$ is shifted to higher temperatures, with increasing frequency, and the corresponding maximum $T$ of the hump ($T_{m}$) is shown in the $\frac{d\, tan\,\delta}{dT}$ curves (Fig.\ref{Fig11.jpg}(b)). Fig.\ref{Fig11.jpg}(b) also shows the frequency-independent peak at 42\,K. Here, we note that the $tan\,\delta$ curves shown in Fig.\ref{Fig11.jpg}(a) are measured using the homemade sample probe in a Quantum Design MPMS system and the data in Fig. \ref{Fig9.jpg} and Fig.\ref{Fig10.jpg} were measured using the Janis cryogenic system. Thus, there is a slight temperature difference in the $T_{MG}$ and $T_{DG} $, but the overall behavior is the same. The frequency-dependent dielectric behavior at low-$T$ is analyzed using the thermally activated 
Arrhenius law 

\begin{equation}
 \tau = \tau_0 \exp\left(\frac{E_a}{k_B T_m}\right)
\end{equation} 

and also Vogel-Fulcher (V-F) relation \cite{Samara,Yang_2023},

\begin{equation}
\tau = \tau_0 \exp\left(\frac{E_a}{k_B (T_m - T_0)}\right)
\end{equation}

Here, $E_{a}$ is the energy barrier to dipole reorientation, $T_{0}$ is the critical temperature at which all the relaxation times diverge, $\tau_{0}$ is the reciprocal of the attempt frequency ($\omega_{0}$) and $T_{m}$ is maximum peak temperature of $\epsilon^{'}$. The $\epsilon^{'}$ vs $T$ curves at 20\,K does not follow the thermally activated Arrhenius law (the nonlinear behavior is shown in Fig.\ref{Fig11.jpg}(c)) and also the fitting leads to a large error in the $\tau_{0}$. The values of the parameters $E_{a}$, $\tau_{0}$, and $T_{0}$\, obtained by a best-fit analysis of the data to the V-F equation are 2.78 ($\pm$0.02)\,meV, 2.47($\pm$0.61)$\times10^{-8} s$, and 10.31($\pm$0.33)\,K, respectively.

The V-F relation is generally used to describe a glassy state in dielectrics and applies to both relaxor ferroelectric (FE) and also dipolar glass\,(DG) \cite{Samara,Yang_2023}. The relaxor FE and DG show similar broad frequency-dependent $\epsilon^{'}$ vs $T$ curves. However, the former can transform into normal FEs under a strong electric field (E), whereas the latter cannot. This is because relaxor FEs contain clustered polar nanoregions (PNRs), randomly interacting with individual electric dipoles of fixed length in DG  \cite{Pirc,Polinger,yin2019,Filipi}. Therefore, to distinguish the LBCO from relaxor FE and DG states, we measured (in cooling) the $\epsilon^{'}$ vs $T$ curves under a DC bias field ($E_{bias}$) of 10 kV/m. The results are shown in Fig.\ref{Fig11.jpg}(c)\,along with the $\epsilon^{'}$ vs $T$ curves measured in the absence of $E_{bias}$. From, Fig.\ref{Fig11.jpg}(c), there no obvious change in the $\epsilon^{'}$ vs $T$ curves under a DC bias field. Further, we have also measured the pyroelectric current at low-$T$ under an external electric field, which shows a broad pyroelectric current peak centered at 55\,K. The peak position shifts along the $T$ axis with changing heating rates. These pyroelectric results are similar to those of YBCO and DBCO cobaltites of the same family \cite{Dhanasekhar_2019}; hence, they are not shown here. The pyroelectric current measurements also show the absence of permanent electric dipoles in LBCO. From these observations, we ruled out the ferroelectric (FE) and relaxor FE features in LBCO, and the low-$T$ dielectric state is assigned to a dipolar glass.

\section {Discussion}

The CHUF, thermal cycling of $M_{ZFC}$, and time-dependent $M$ studies clearly show glass-like dynamics at low-$T$, where the glassy phase arises from the kinetic arrest of the supercooled magnetic phase. The bulk $M$ measurements are sufficiently sensitive to detect the phase fractions if the kinetic arrest is related to long-range magnetic order, involving the ferromagnetic and anti-ferromagnetic ordered phases \cite{Pal2023}. For, LBCO, neutron studies show short-range magnetic correlations and the absence of long-range magnetic order. Thus, it is challenging to identify the phase fractions at low-$T$ in LBCO from the $M$ studies. However, in previous X-ray, and neutron studies, along with cooling rate-dependent $M$ studies, the $M$, peak at 48 K was assigned to the$Cc$  phase \cite{Avci}. At low-$T$, in the R-114 compounds, the $Pbn2_{1}$ phase is more stable and exhibits AFM (or FiM) ordering, whereas the $Cc$  phase is observed only in LBCO and is paramagnetic (PM) in nature at high-$T$. At 110\,K, the $Cc$ phase is converted to $Pbn2_{1}$, and at low-$T$ the phase fractions strongly depend on the cooling rates \cite{Avci}. In the present experiments, we used the 3\,K/min in the $M$ studies, and 2\,K/min in the $\epsilon^{'}$ studies, which indicates LBCO contains both $Pbn2_{1}$, $Cc$ phases qualitatively. However, as indicated above, the exact phase fractions cannot be predicted. If the high-$T$, PM $Cc$ phase is kinetically arrested at low-$T$, without showing any long-range magnetic order, then one would expect the $M$ associated with this phase to decrease linearly without showing any peak, i.e., pure PM behavior. On the other hand, the $M$ investigations clearly display an AFM-like peak at 48\,K suggesting that the $Pbn2_{1}$ phase fraction contributes significantly to the kinetic arrest and might be produce a peak at 48\,K, during heating. This can be further understood from the thermal cycling of $M_{ZFC}$ (Fig.\ref{Fig7.jpg}), where the peak at 48 K, becomes weak when we perform the $T$ cycling in 60 K - 100 K and also at the low-$T$, $M$ shows linear decrease. We note that in the thermal cycling of $M_{ZFC}$, a heating and cooling rate of 3 K/min was used and it was observed that the$M_{ZFC}$ peak at 48 K becomes weaker. A similar behavior is also reported in \cite{Avci}, where the $M$ was measured using slower cooling rates ($<$ 1 K/min).

We now present additional evidence suggesting that the peak near 48\,K might be associated with the $Pbn2_{1}$ phase by comparing the $\epsilon^{'}$ properties of LBCO with the reported $\epsilon^{'}$ studies of other R-114 cobaltites. The crystal structure of CBCO is orthorhombic $Pbn2_{1}$ (300\,K to 4\,K; non-centrosymmetric) and exhibits long -range FiM ordering at low-$T$ ($T_{c}$ $\approx$ 60\,K). The $\epsilon^{'}$ studies on CBCO at low-$T$ show a frequency-independent peak at magnetic ordering $T$. For $T$ $<$ $T_{c}$, $\epsilon^{'}$ of CBCO decreases with decreasing $T$, without showing any relaxation features \cite{CDS2016,CDS2017,Singh,Caignaert}. However, the dielectric studies on the YBCO and DBCO cobaltites show the absence of the $\epsilon^{'}$ peak at low-$T$ and is also constant for $T$ $<$ $30\, K$ \cite{Dhanasekhar_2019}. Here, we note that the magnetic ground states of these YBCO and DBCO cobaltites are different, DBCO shows spin glass-like behavior along with short-range magnetic correlations, whereas YBCO shows both long-range and short-range magnetic features below 100\,K \cite{Manuel2009,Khalyavin,Dhanasekhar_2019}. Based on these  $\epsilon^{'}$ comparisons, we attribute the frequency-independent peak of LBCO  at $T$ = 43 K to the  $Pbn2_{1}$ phase. Further, as discussed above, the kinetically arrested PM $Cc$ phase at low-$T$ in LBCO would likely show dipolar glassy features in the dielectric studies.

\section {Conclusion}

In conclusion, we report experimental evidence of magnetic and dielectric glassy states in alternating kagome and triangular lattice LuBaCo$_{4}$O$_{7}$ cobaltite. AC magnetization studies show an absence of the conventional spin glass state, whereas the various detailed DC $M$ studies show the magnetic glassy state at low-$T$, arising from the kinetic arrest of the first-order structural phases. Thermal cycling measurements show that broad thermal hysteresis plays a significant role on the arrested low-$T$ magnetic phases. The dielectric features of the LBCO are strongly coupled with structural phase transitions and also show strong signatures of kinetically arrested phases. Recent studies have established that the phase fractions at low temperature in LBCO vary significantly with the Y doping at Lu sites \cite{Sahinbay}. Thus, studying Lu$_{1-x}$Y$_{x}$BaCo$_{4}$O$_{7}$, systematically may shed more light on the origin of the kinetic arrest-driven magnetic and dielectric glassy features and the role of phase coexistence in these systems.

\section*{Acknowledgment}
This work is supported by the Science and Engineering Research Board-National Post-Doctoral Fellowship (PDF/2021/002536) funding. This study is supported by the Ministry of Science and Technology, Taiwan, under Grants No. NSTC-112-2112-M-110-018 and NSTC-113-2112-M-110-006. The author group from IIT Bombay acknowledges the support of central measurement facilities (HR-XRD, SQUID MPMS) at their institution. The authors from IIT Kharagpur acknowledge support of central SQUID magnetometer facilities. 

\appendix
\section{impedance studies}

\begin{figure*}
\includegraphics[width=0.8\textwidth]{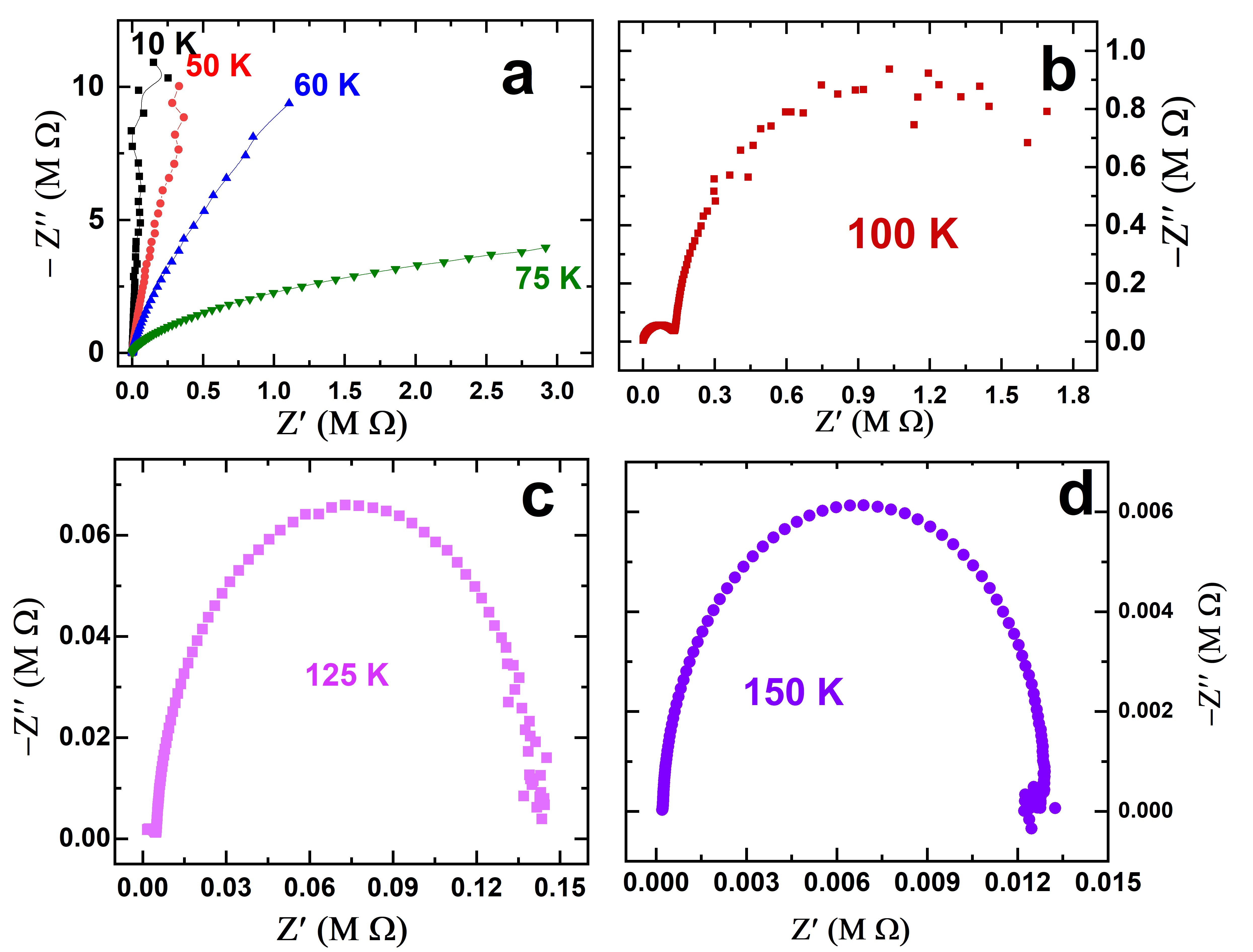}
\centering
\caption { The (a)-(d) shows the $T$ and frequency variation of the real ($Z'$) and imaginary ($Z''$) parts of the electrical impedance of LBCO}.
\label{Fig12.jpg}	
\end{figure*}

The variation of the real ($Z'$) and imaginary ($Z''$) parts (known as the Cole-Cole plot or the Nyquist plot) of the electrical impedance with frequency at different temperatures is shown in Fig.\ref{Fig12.jpg}. The appearance of clear semicircle arcs at T $>$ 75 K indicates the presence of two types of relaxation phenomena, such as intrinsic grain and grain boundary components. As the $T$  decreases, the semi-circle arcs disappear within the measured frequency range. These results show that at high-$T$ both grain and grain boundary effects coexist, whereas at low-$T$ only contributions of intrinsic grains present in the LBCO. The formation of internal barriers at the grain boundaries probably leads to a large $\epsilon^{'}$ at high-$T$.

\bibliographystyle{apsrev4-1}
\bibliography{bib}
\end{document}